\documentstyle[PASJadd,psbox]{PASJ95}
\newcommand{\lett}{}

\begin{document}
\title{Discovery of 172-s Pulsations from a Be/X-ray Binary Candidate 
AX~J0051.6$-$7311 in the SMC with ASCA
}
\author{Jun {\sc Yokogawa},$^1$
Ken'ichi {\sc Torii},$^2$
Kensuke {\sc Imanishi},$^1$
and Katsuji {\sc Koyama}$^1$\thanks{CREST, 
Japan Science and Technology Corporation (JST), 
4-1-8 Honmachi, Kawaguchi, Saitama, 332-0012.} \\ [12pt]
$^1$ {\it Department of Physics, Graduate School of Science, Kyoto University, 
Sakyo-ku, Kyoto, 606-8502} \\
{\it E-mail(JY): jun@cr.scphys.kyoto-u.ac.jp} \\
$^2$ {\it National Space Development Agency of Japan, 
2-1-1 Sengen, Tsukuba, Ibaraki, 305-8505}}

\abst{The results from three ASCA observations of 
AX~J0051.6$-$7311 = RX~J0051.9$-$7311 
are reported. 
Coherent pulsations with a barycentric period of $172.40 \pm 0.03$~s 
were discovered in the third observation, 
with an exceptionally long exposure time of $\sim 177$~ks. 
The X-ray spectrum was found to remain unchanged through these 
observations, with a photon index of $\sim 0.9$. 
Energy-resolved pulse profiles in the third observation reveal that 
the pulsations are mostly due to photons with an energy 
above $\sim 2$~keV. 
Archival data of ROSAT and Einstein indicate 
that AX~J0051.6$-$7311 exhibits a flux variation having a factor $\gtsim 20$. 
}

\kword{pulsars: individual (AX~J0051.6$-$7311) --- 
stars: emission-line, Be --- stars: neutron --- X-rays: stars}

\maketitle
\thispagestyle{headings}

%section 1
\section{Introduction}
Candidates for X-ray binary pulsars (XBPs) 
in the Small Magellanic Cloud (SMC) 
have been identified in two ways. 
Yokogawa et al.\ (2000a) observed the main body and 
the eastern wing of the SMC with ASCA, 
and made systematic analyses on the detected X-ray sources. 
They found that XBPs and thermal supernova remnants
are clearly separated by the spectral hardness ratio. 
Eight candidates for XBPs were found from this classification. 
Probably, no detection of coherent pulsations from these candidates 
is due to the limited photon statistics. 
On the other hand, Cowley et al.\ (1997) and Schmidtke et al.\ (1999) 
made a systematic search of optical counterparts for X-ray sources. 
They determined accurate X-ray positions by ROSAT/HRI and 
made optical photometry and spectroscopy 
within the X-ray error circles. 
They identified five X-ray sources with a Be star companion, 
which are good candidates for XBPs.

The most promising and direct way to establish an XBP  
is to search for coherent pulsations from these candidates. 
ASCA observations are suitable for a pulsation search 
because of high sensitivity in the hard X-ray band. 
In fact, we searched for coherent pulsations with ASCA from 
RX~J0050.8$-$7316, a binary with 
a Be star companion (Cowley et al.\ 1997), 
and discovered 323-s pulsations (Imanishi et al.\ 1999). 
However, XBPs with a Be star companion are usually in the low-luminosity state 
of $\ltsim 10^{36}$ erg~s$^{-1}$ 
(Bildsten et al.\ 1997; Stella et al.\ 1986), 
or a flux of $\ltsim 2 \times 10^{-12}$ erg~s$^{-1}$~cm$^{-2}$ 
at a distance of the SMC (60 kpc is assumed in this paper; van den Bergh 2000). 
It is difficult to detect coherent pulsations from 
such a low-flux source for a nominal ASCA exposure time of $\sim 40$ ks. 
Therefore, 
we made an ASCA observation with an exceptionally long 
exposure time of $\sim 177$ ks. 
Within the observed region is located 
AX~J0051.6$-$7311 = RX~J0051.9$-$7311,  
which is regarded as an XBP candidate (Yokogawa et al.\ 2000a), 
and has also been identified with a Be star companion (Cowley et al.\ 1997), 
and is this the most promising target for a pulsation search. 

In this letter, we report on the discovery of coherent pulsations 
from AX~J0051.6$-$7311 (Torii et al.\ 2000) 
with a very long exposure observation. 
We also report a long-term flux history based on investigating 
the archival data of ROSAT and Einstein.

%section 2
\section{ASCA Observations and Data Reduction}
Three ASCA observations (hereafter obs.\ A1--A3) 
have covered the position of AX~J0051.6$-$7311. 
The time spans of these observations were 
50765.179--50766.285 (A1), 
51309.604--51310.682 (A2), and 
51645.986--51651.490 (A3) in units of Modified Julian Day (MJD).

ASCA carries four XRTs (X-ray Telescopes, Serlemitsos et al.\ 1995) 
with two GISs (Gas Imaging Spectrometers, Ohashi et al.\ 1996) and 
two SISs (Solid-state Imaging Spectrometers, Burke et al.\ 1994)
on each focal plane.
In this paper, we do not refer to SIS, because 
AX~J0051.6$-$7311 was outside of the SIS field of view 
in all observations. 
The GIS was operated in the normal PH mode 
in all observations, providing
a time resolution of 0.0625~s (high bit rate) or 0.5~s (medium bit rate).
We screened the GIS data 
by rejecting any data obtained in the South Atlantic Anomaly 
or low cut-off rigidity regions ($<4$~GV). 
Data obtained when the satellite's elevation angle was lower than 
$5^\circ$ were also rejected. 
Particle events were removed by a rise-time discrimination method.
After screening, the total available exposure times 
in obs.\ A1, A2, and A3 were
$\sim43$~ks, $\sim 41$~ks, 
and $\sim177$~ks, respectively.
In the following analyses, we use the data from 
GIS-2 and GIS-3 simultaneously.

%section 3
\section{Results}
%section 3.1
\subsection{Source Identification}
In obs.\ A3, a dim source was detected at 
$\sim 13'$ from the center of the field of view (FOV). 
Its position was determined to be 
(00$^{\rm h}$51$^{\rm m}$38$^{\rm s}$, $-$73$^\circ$11$'$01$''$)
with an error of $\sim 1'\hspace{-2pt}.5$; 
thus, we designate it as AX~J0051.6$-$7311.
This source was also detected in obs.\ A1 and A2 
at $\sim 21'$ from the center of FOV, 
with a positional error of $\sim 1'\hspace{-2pt}.5$--$2'$. 

In order to determine an X-ray counterpart, 
we investigated several source catalogs of the Einstein and the 
ROSAT satellites. 
An Einstein source No.\ 25 in Wang and Wu (1992) and 
a ROSAT source RX~J0051.9$-$7311 (Cowley et al.\ 1997; Haberl et al.\ 2000)
were found within the error region of ASCA; 
hence, these would be the same source.  
Cowley et al.\ (1997) determined the most accurate position 
with ROSAT HRI to be (00$^{\rm h}$51$^{\rm m}$51$^{\rm s}$\hspace{-2pt}.4, 
$-$73$^\circ$10$'$38$''$), with a $5''$ error radius. 
Figure 1 shows a ROSAT HRI image around RX~J0051.9$-$7311 
with the error regions of Einstein and ASCA.  

%Place fig. 1 here.
\begin{figure}[th]
\hspace*{8mm} \psbox[xsize=0.4\textwidth]{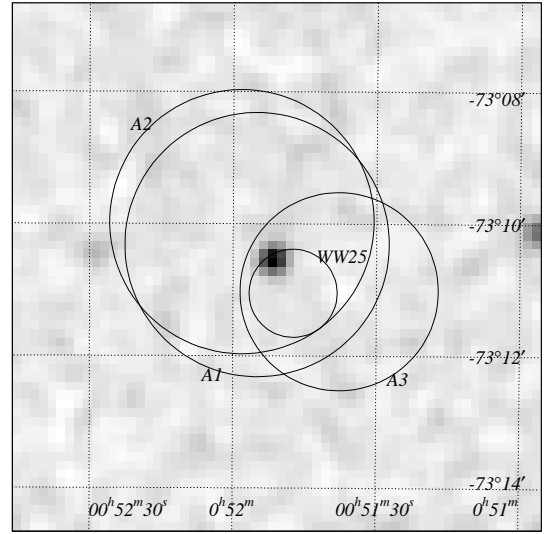}
\caption{ROSAT HRI image (sequence ID. rh600811n00) around RX~J0051.9$-$7311. 
The error circles of AX~J0051.6$-$7311 
(A1--A3) and Einstein source No.\ 25 (WW25) are overlaid.}
\end{figure}

%section 3.2
\subsection{Timing Analyses}
In each observation, 
we collected source photons from a circle 
centered on AX~J0051.6$-$7311. 
The radius of the circle was set to be $2'$ 
in order to avoid contamination from nearby sources 
located $\gtsim 4'$ away. 
After a barycentric correction on the photon arrival times, 
we performed an FFT (Fast Fourier Transformation) analysis.
Only in the power spectrum of obs.\ A3, 
we detected an evident peak at a frequency of 0.0058~Hz. 
The FFT analysis was performed for various energy ranges  
to maximize the S/N ratio; finally, 
the maximum power of 59.1 was obtained from the photons in 2.6--8.6~keV
(figure 2). 
The probability to detect such a strong power at any frequency 
from random events 
was estimated to be $\sim 8 \times 10^{-8}$, 
and hence the pulse detection is highly significant. 
We then performed an epoch folding search, assuming no period change 
during obs.\ A3.  
The barycentric period was determined to be $P=172.40 \pm 0.03$~s.

\begin{figure}
\hspace*{12mm}\psbox[xsize=0.4\textwidth]{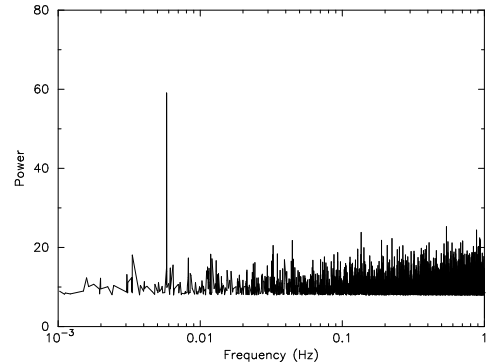}
\caption{Power spectrum in obs.\ A3. Data points with power less than 8 are omitted. 
An evident peak is detected at $\sim 0.0058$~Hz.}
\end{figure}

Figure 3 shows the pulse profiles in the low- and high-energy bands 
(0.7--2.6~keV and 2.6--8.6~keV) 
and their intensity ratio. 
We clearly found the energy dependence of the pulse profiles; 
high energy photons are responsible for most of the pulsations. 
The pulsed fraction, defined as (pulsed flux)/(total flux) without 
background, is $\sim 60$\% in 2.6--8.6~keV. 
To investigate a period change during obs.\ A3, 
we divided the observation span in half and performed epoch folding
searches on both of the segments; however, no significant period change was 
detected. 

%Place fig. 2 here.
%Place fig. 3 here.

We also examined the aperiodic intensity variation during 
each observation, 
by binning  the light curves with  
several time scales from $\sim$~second to $\sim$~hour. 
Neither a large flux variation nor any burst-like activity was found
in each observation. 
Weak evidence for a random variation of a factor $\sim 2$ 
was found in obs.\ A1 and A3 with a time-scale of 5000--10000~s.

%section 3.3
\subsection{Spectral Analyses}
X-ray spectra were extracted from the same circles 
used in the timing analyses,  
while background spectra were 
from source-free areas near to the source regions. 

Figure 4 shows the background-subtracted and 
phase-averaged spectrum in obs.\ A3, which has the best statistics.
Since no emission line was found in the spectra of obs.\ A1--A3, 
we fitted each spectrum to an absorbed power-law model. 
The spectral parameters (photon index $\Gamma$ and column density $N_{\rm H}$) 
were found to be consistent within the statistical error 
throughout these observations. 
Therefore, in order to obtain better statistics, we assumed that $\Gamma$ and $N_{\rm H}$ are 
constant in these observations, and simultaneously fitted 
the three spectra again. 
We obtained an acceptable fit 
with the best-fit parameters 
given in table 1. 

For obs.\ A3, we also extracted phase-resolved spectra 
from phases 0--0.5 (``on-pulse'') and 0.5--1 (``off-pulse'') (see figure 3). 
We fitted the spectra to the same model as given in table 1. 
We found that $\Gamma$ is much smaller in the on-pulse phase,
which is consistent with the energy-resolved pulse shape 
(figure 3).
We also found a hint of soft excess below $\sim 1$~keV 
only in the on-pulse spectrum. 
If we added a blackbody component to the model, 
the temperature and unabsorbed luminosity were determined to be 0.13~keV 
and $2\times10^{35}$ erg~s$^{-1}$~cm$^{-2}$ (0.7--10.0~keV), respectively, 
with smaller $\chi^2$ of 35.03 for 27 d.o.f..

%section 4
\section{Discussion}
The position of AX~J0051.6$-$7311 has previously been covered by 
14 observations of Einstein and ROSAT. 
We used the archival data of these observations 
to investigate a long-term flux variation of AX~J0051.6$-$7311. 
After background subtraction, 
we derived the count rate of AX~J0051.6$-$7311 in each observation. 
We then converted the count rate to the flux with the {\tt PIMMS} software, 
assuming that $\Gamma$ and $N_{\rm H}$ are unchanged from 
the best-fit values \ for \  the  \ phase-averaged \ spectra \ (table 1). \ 
The 

%\begin{table*}[b]
{\small %$B$3$l$O=PHG;~$N$_(B
%\begin{center}
%\end{center}
\vspace{8pt}
\begin{tabular*}{\textwidth}{@{\hspace{\tabcolsep}
\extracolsep{\fill}}p{14pc}cccccc}
\multicolumn{6}{c}{Table~1.\hspace{4pt}Results of the spectral fitting.}\\[9pt]
\hline\hline\\ [-6pt]
Obs.\ ID.   &$\Gamma$       & $N_{\rm H}$           & Flux                & Luminosity     & $\chi^2$/d.o.f. \\
            &               & (cm$^{-2}$)      & (erg s$^{-1}$ cm$^{-2}$) & (erg s$^{-1}$) & \\[4pt]\hline\\[-6pt]
\multicolumn{5}{l}{--- Phase-averaged ---} \\
 A1\dotfill & *             & *                     &$6.8\times 10^{-13}$ & $2.9\times 10^{35}$  & *\\
 A2\dotfill & *             & *                     &$1.5\times 10^{-12}$ & $6.4\times 10^{35}$  & *\\
 A3\dotfill & 0.9 (0.8--1.0)&0 ($<8\times 10^{20}$) &$1.3\times 10^{-12}$ & $5.6\times 10^{35}$  & 61.5/69 \\[4pt]
\hline\\[-6pt]
\multicolumn{5}{l}{--- Phase-resolved ---} \\
 A3: on-pulse\dotfill  &0.6 (0.4--0.8)& 0 ($<1\times 10^{21}$) &$1.7\times 10^{-12}$ &$7.3\times 10^{35}$ & 43.3/29  \\
 A3: off-pulse\dotfill &1.4 (1.1--1.7)& 0 ($<3\times 10^{21}$) &$8.5\times 10^{-13}$ &$3.7\times 10^{35}$ & 28.9/24  \\[4pt]
\hline\\[-8pt]
%\vspace{6pt}
\multicolumn{6}{l}{Note --- Parentheses indicate 90\% confidence limits. Flux and luminosity are derived from the energy band of 0.7--10.0~keV.}\\
\multicolumn{6}{l}{$*$ $\Gamma$ and $N_{\rm H}$ are linked between the observations. }
\end{tabular*}
}

\begin{figure}[t]
\hspace*{8mm}\psbox[xsize=0.4\textwidth]{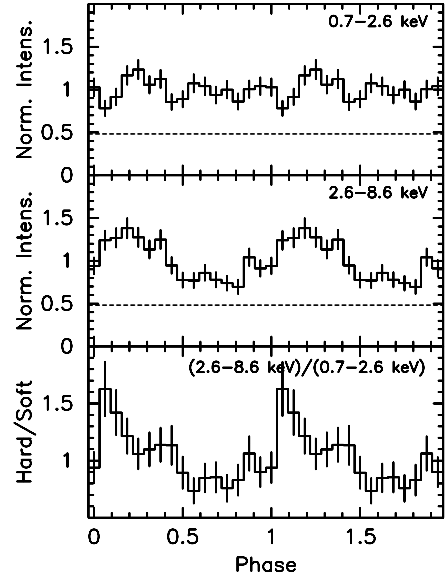}
\caption{Pulse profiles in the 0.7--2.6~keV (upper panel) and 
2.6--8.6~keV (middle panel) bands, and the intensity ratio between them 
(lower panel). Background levels are indicated by broken lines.}
\end{figure}

%Place fig. 4 here.
%Place table 1 here.

\begin{figure}
\hspace*{8mm}\psbox[xsize=0.4\textwidth]{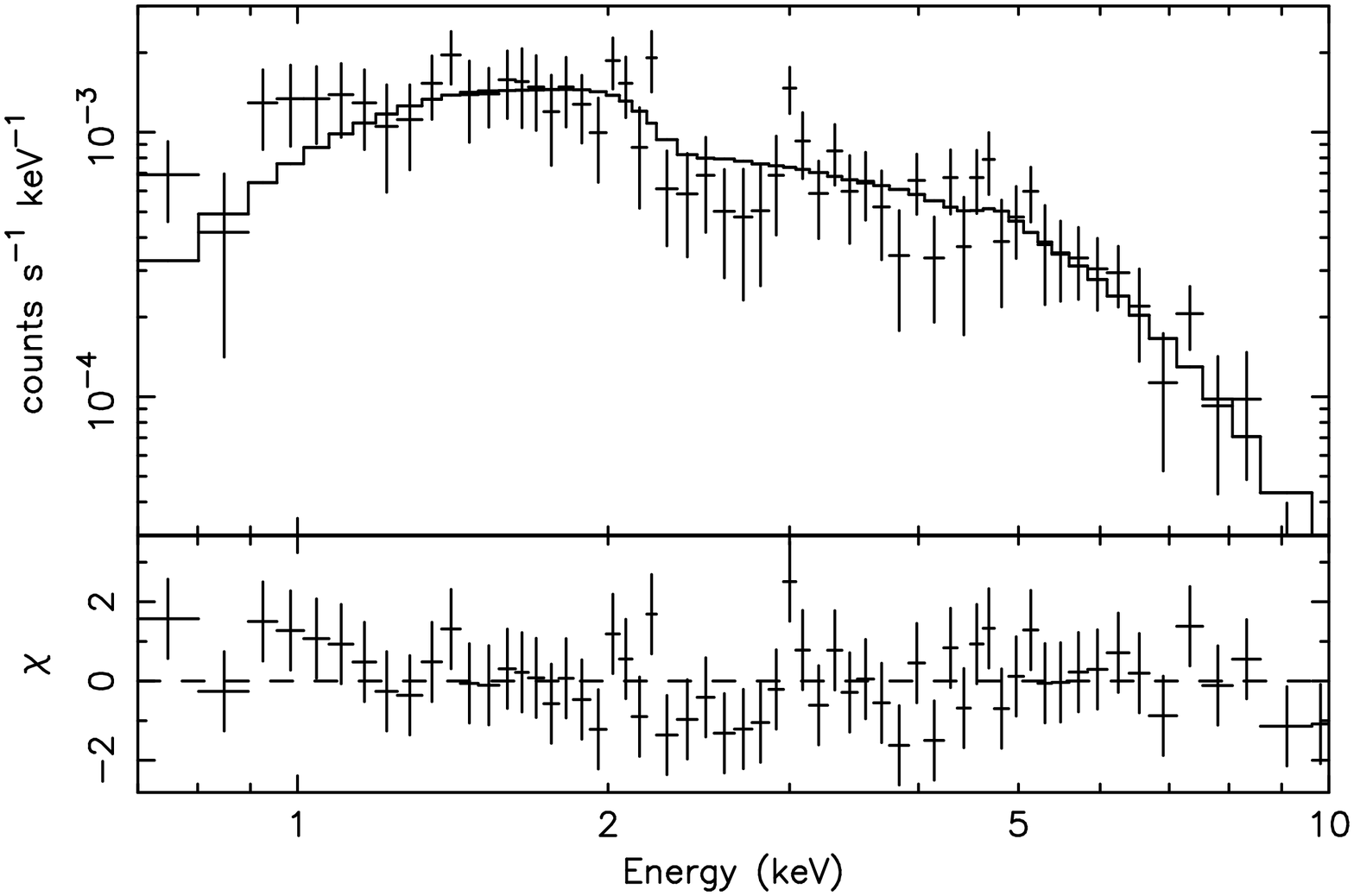}
\caption{Background-subtracted and phase-averaged spectrum from GIS2+3 in obs.\ A3. 
Crosses and a solid line indicate data points and 
the best-fit model, respectively. 
The best fit model was obtained by the spectral fitting 
using all the data from obs.\ A1, A2, and A3, simultaneously.
}
\end{figure}

\newpage
\ 
\newpage

%
%\begin{table*}[t]
{\small
\begin{center}
Table~2.\hspace{4pt}Flux variability of AX~J0051.6$-$7311. \\[9pt]
\end{center}
%\vspace{3pt}
\begin{tabular*}{0.48\textwidth}{@{\hspace{\tabcolsep}
\extracolsep{\fill}}p{2.5pc}ccc}
\hline\hline\\ [-6pt]
Obs.ID.&Date$^*$  &  Instrument  &  Flux$^\dagger$         \\
       &(MJD)     &              &  (erg~s$^{-1}$~cm$^{-2}$)  \\[4pt]\hline\\[-6pt]
%E1\dotfill& 43995.264 & Einstein/IPC & $5.2\times10^{-14}$ \\
E1\dotfill& 43995.264 & Einstein/IPC &$<1.4\times10^{-13}$ \\
E2\dotfill& 44189.468 & Einstein/IPC & $3.4\times10^{-13}$ \\
E3\dotfill& 44345.640 & Einstein/IPC & $1.3\times10^{-13}$ \\
R1\dotfill& 48550.301 & ROSAT/PSPC   & $8.5\times10^{-13}$ \\
R2\dotfill& 48732.665 & ROSAT/PSPC   & $7.7\times10^{-13}$ \\
R3\dotfill& 48963.036 & ROSAT/PSPC   &$<3.8\times10^{-14}$ \\
%R4\dotfill& 49093.121 & ROSAT/PSPC   & $3.4\times10^{-14}$ \\
R4\dotfill& 49093.121 & ROSAT/PSPC   &$<8.6\times10^{-14}$ \\
R5\dotfill& 49118.070 & ROSAT/PSPC   & $3.4\times10^{-13}$ \\
R6\dotfill& 49298.538 & ROSAT/PSPC   & $3.4\times10^{-13}$ \\
R7\dotfill& 49321.271 & ROSAT/PSPC   & $6.7\times10^{-13}$ \\
R8\dotfill& 49652.016 & ROSAT/HRI    & $8.9\times10^{-13}$ \\
R9\dotfill& 49864.713 & ROSAT/HRI    & $1.2\times10^{-12}$ \\
R10\dotfill&50055.454 & ROSAT/HRI    & $1.8\times10^{-12}$ \\
R11\dotfill&50412.933 & ROSAT/HRI    & $6.5\times10^{-13}$ \\
A1\dotfill& 50765.732 & ASCA/GIS     & $6.8\times10^{-13}$ \\
A2\dotfill& 51310.143 & ASCA/GIS     & $1.5\times10^{-12}$ \\
A3\dotfill& 51648.738 & ASCA/GIS     & $1.3\times10^{-12}$ \\[4pt]
\hline\\[-8pt]
\multicolumn{4}{l}{$*$ Middle of the observations. }\\
\multicolumn{4}{l}{$\dagger$ In the 0.7--10.0~keV band. }
\end{tabular*}
%\end{table*}
%
}

\ 

\noindent
results are given in table 2. 
We found that AX~J0051.6$-$7311 may have been brightest 
in the ASCA observations. 
The largest flux difference having a factor $\gtsim 50$
is found between obs.\ R3 and obs.\ R10, 
although there may be some systematic error between different instruments 
(PSPC and HRI). 
However, even if we restrict the flux comparison within the same instruments, 
we find a flux difference having a factor $\gtsim 20$ 
between obs.\ R1 and obs.\ R3 (PSPC). 
Such a large flux variation, together with the long period (172~s), 
hard spectrum ($\Gamma=0.9$), and a Be star counterpart, 
suggests that AX~J0051.6$-$7311 should be an XBP with a Be star companion. 

%Place table 2 here.

%\ 

A small bump is found in the pulse profile of the lower energy band 
(figure 3; phase 0.1--0.4). 
The soft excess detected in the on-pulse phase 
implies existence of a pulsating soft component found in some XBPs 
(e.g., Her X-1, Endo et al.\ 2000; 
LMC X-4, Woo et al.\ 1996; 
XTE J0111.2$-$7317, Yokogawa et al.\ 2000b), 
although better statistics are necessary to lead to a rigid conclusion; 
a simple power-law model is also accepted 
for the on-pulse spectrum 
within 95\% confidence level (see table 1).

Coherent pulsations were detected only from the data of obs.\ A3. 
We suggest that this is simply due to the much improved 
statistics in obs.\ A3, with the unusually long exposure time 
($\sim 177$~ks).
Indeed, when we divided obs.\ A3 into four segments, 
each having an exposure time of $\sim 40$~ks, 
we could not detect significant coherent pulsations 
from any of these segments 
neither by the FFT analysis nor 
by the epoch folding search.
We also could not detect coherent pulsations from obs.\ R2,
which has the best statistics among the ROSAT/Einstein observations.
Coherent pulsations have been detected from 
all of the hard X-ray sources in the SMC 
with a luminosity of $\gtsim 10^{36}$ erg~s$^{-1}$~cm$^{-2}$ 
(Yokogawa et al.\ 2000a), 
within a typical ASCA exposure of $\sim 40$~ks. 
Pulsations from fainter hard X-ray sources, 
AX~J0051.6$-$7311 and 1SAX~J0103.2$-$7209, 
with luminosities of $\sim 5\times10^{35}$ erg~s$^{-1}$~cm$^{-2}$, 
were also detected with much longer exposures 
(this work and Yokogawa et al.\ 2000a).
These facts strongly suggest that 
a significant fraction of 
the unidentified hard sources in Yokogawa et al.\ (2000a) 
with a luminosity of $\ltsim 10^{36}$ erg~s$^{-1}$~cm$^{-2}$ 
are XBPs. 
Therefore, a long observation like obs.\ A3 is highly required to 
extend the pulsar population study to the low luminosity side. 
\par
\vspace{1pc}\par
The Einstein and ROSAT data were obtained through the High Energy
Astrophysics Science Archive Research Center Online Service,
provided by the NASA/Goddard Space Flight Center. 
J.Y.\ is supported by JSPS Research Fellowship for Young Scientists.

%\clearpage
\section*{References} 
\small
\re
 Bildsten L., Chakrabarty D., Chiu J., Finger M.H., Koh D.T., 
 Nelson R.W., Prince T.A., Rubin B.C.\ et al.\ 1997, ApJS 113, 367
\re
 Burke B.E., Mountain R.W., Daniels P.J., Dolat V.S., 
 Cooper M.J.\ 1994, IEEE Trans.\ Nucl.\ Sci.\ 41, 375
\re
 Cowley A.P., Schmidtke P.C., McGrath T.K., Ponder A.L., 
 Fertig M.R., Hutchings J.B., Crampton D.\ 1997, PASP 109, 21
\re
 Endo T., Nagase F., Mihara T.\ 2000, PASJ 52, 223
\re
 Haberl F., Filipovi\'{c} M.D., Pietsch W., Kahabka P.\ 
 2000, A\&AS 142, 41
\re
 Imanishi K., Yokogawa J., Tsujimoto M., Koyama K.\ 1999, PASJ 51, L15 
\re
 Ohashi T., Ebisawa K., Fukazawa Y., Hiyoshi K., 
 Horii M., Ikebe Y., Ikeda H., Inoue H.\ 
 et al.\ 1996, PASJ 48, 157
\re
 Schmidtke P.C., Cowley A.P., Crane J.D., Taylor V.A., McGrath T.K., 
 Hutchings J.B., Crampton D.\ 1999, AJ 117, 927
\re
 Serlemitsos P.J., Jalota L., Soong Y., Kunieda H., Tawara Y., 
 Tsusaka Y., Suzuki H., Sakima Y.\ 
 et al.\ 1995, PASJ  47, 105
\re
 Stella L., White N.E., Rosner R.\ 1986, ApJ 308, 669
\re
 Torii, K., Yokogawa, J., Imanishi, K., Koyama, K.\ 2000, IAU Circ.\ 7428
\re
 van den Bergh S.\ 2000, PASP 112, 529
\re
 Wang Q., Wu X.\ 1992, ApJS 78, 391
\re
 Woo J.W., Clark G.W., Levine A.M., Corbet R.H.D., Nagase F.\ 1996, ApJ
 467, 811
\re
 Yokogawa J., Imanishi K., Tsujimoto M., Nishiuchi M., Koyama K., 
 Nagase F., Corbet R.H.D.\ 2000a, ApJS in press
\re
 Yokogawa J., Paul B., Ozaki M., Nagase F., Chakrabarty D., 
 Takeshima T.\ 2000b, ApJ 539, 191

\end{document}